\documentclass{article}

\usepackage[english]{babel}
\usepackage[misc]{ifsym} 
\usepackage{biblatex}
\usepackage{indentfirst}
\usepackage[letterpaper,top=2cm,bottom=2cm,left=3cm,right=3cm,marginparwidth=1.75cm]{geometry}

\usepackage{amsmath}
\usepackage{graphicx}
\usepackage{pdfpages}
\usepackage{authblk}
\usepackage[colorlinks=true, allcolors=blue]{hyperref}

\date{}
\title{\textbf{A Color Image Steganography Based on Frequency Sub-band Selection}}

\author[1]{Hai Su}
\author[1]{Shan Yang}
\author[1]{Shuqing Zhang}
\author[1]{{Songsen Yu} \thanks{Songsen Yu: yss8109@163.com}}

\affil[]{School of Software,South China Normal University,528200,Guangdong,China}

\begin{document}

\maketitle
\begin{abstract}
Color image steganography based on deep learning is the art of hiding information in the color image. Among them, image hiding steganography(hiding image with image) has attracted much attention in recent years because of its great steganographic capacity. However, images generated by image hiding steganography may show some obvious color distortion or artificial texture traces. We propose a color image steganographic model based on frequency sub-band selection to solve the above problems. Firstly, we discuss the relationship between the characteristics of different color spaces/frequency sub-bands and the generated image quality.   Then, we select the B channel of the RGB image as the embedding channel and the high-frequency sub-band as the embedding domain. DWT(discrete wavelet transformation) transforms B channel information and secret gray image into frequency domain information, and then the secret image is embedded and extracted in the frequency domain. Comprehensive experiments demonstrate that images generated by our model have better image quality, and the imperceptibility is significantly increased.
\end{abstract}

\section{Introduction}

Color image steganography is one of the essential branches of information hiding. It is widely used in the military, medicine, intellectual property protection, and other fields. Through human visual characteristics, color image steganography can hide secret information (usually bitstream or text) into a color image. We call the original color image a cover image, and the modified color cover image a stego image. The stego image is almost the same as the cover image, so it is not easy to arouse the suspicion of the third party and shows good imperceptibility. In recent years, the image hiding steganography based on deep learning has emerged. Image hiding steganography can hide the image as secret information into a cover image. This kind of steganography improves the steganographic capacity and greatly expands the application scope of steganography.

\begin{figure}[t]
\centering
\includegraphics[width=1.0\textwidth]{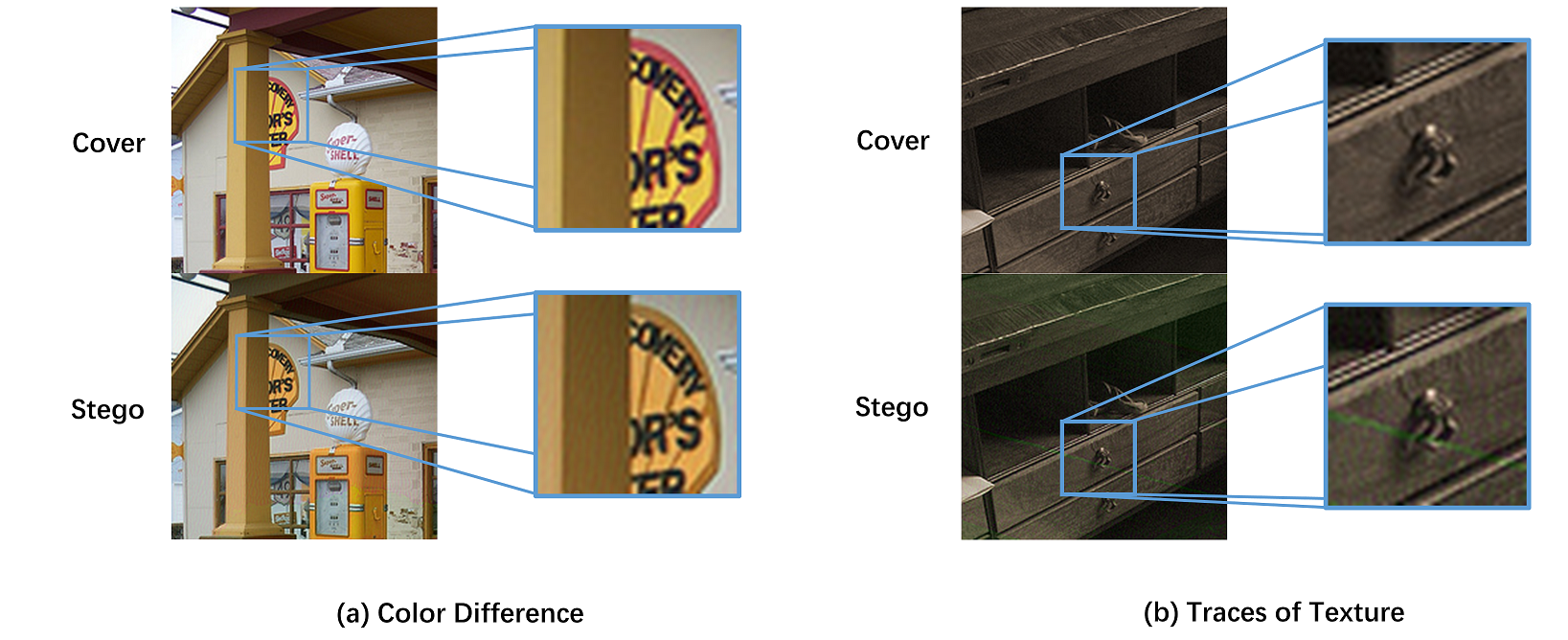}
\caption{\label{fig:1-1}We mark and enlarge the color distortion (a) and artificial texture traces (b) area in stego images generated by \cite{r3}.}
\end{figure}

\begin{figure}[t]
\centering
\includegraphics[width=1.0\textwidth]{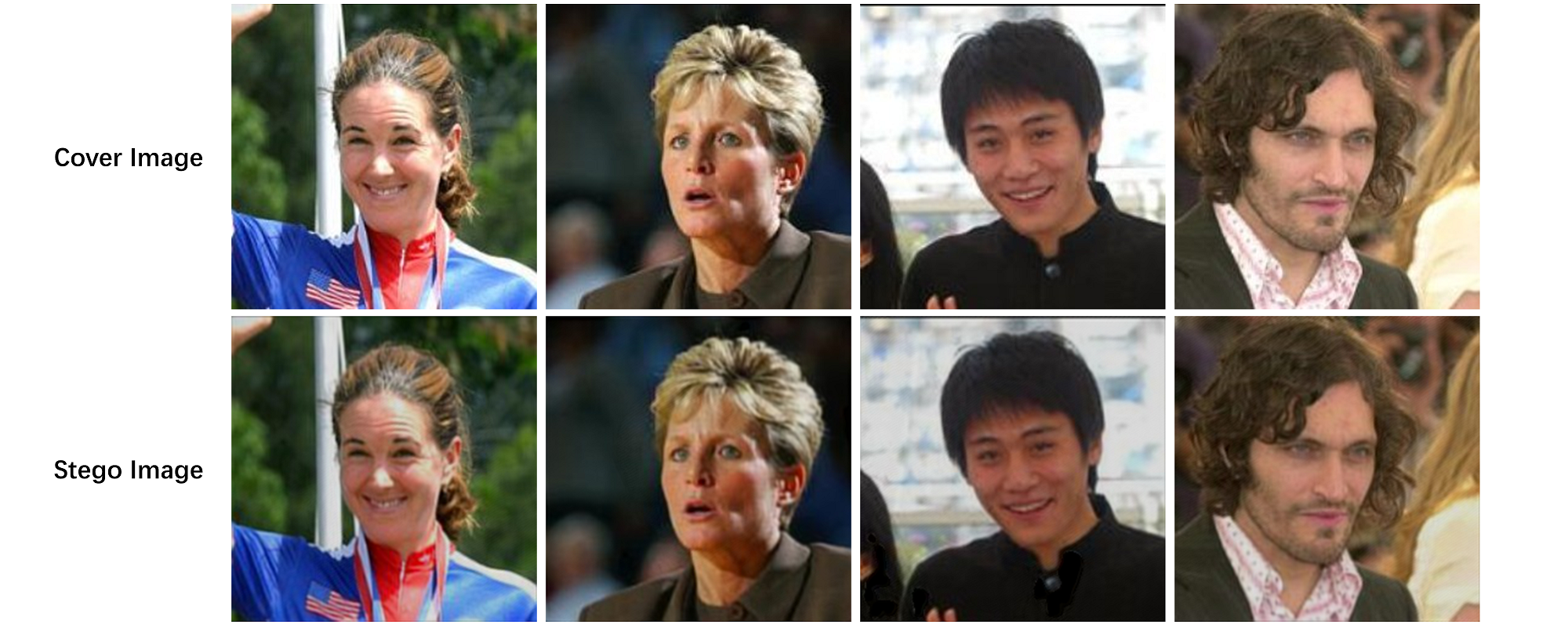}
\caption{\label{fig:1-2}Four groups of stego images generated by the model of \cite{r4}. This model uses the characteristics of YUV color space to improve the quality of stego images. There is only a slight difference in brightness between stego images and original images.}
\end{figure}

Nowadays, image hiding steganography based on deep learning is a popular research direction in steganography. Image hiding steganography should have a large steganographic capacity and good imperceptibility to obtain a better application effect. However, there is a strong negative correlation between capacity and imperceptibility. The increase of steganographic capacity usually leads to a decrease in imperceptibility\cite{r1}.Therefore, it is difficult for the current image hiding steganography based on deep learning to take into account these two characteristics simultaneously\cite{r2}. As shown in Figure \ref{fig:1-1} , when the secret information is the same size image, the stego image may generate visible color distortion and artificial texture traces, which do not meet the imperceptibility requirement of image steganography. Therefore, currently controlling the steganographic traces in the invisible range of human eyes is the point of image hiding steganography. As shown in Figure \ref{fig:1-2}, some researchers tried to use the characteristics of color spaces and embedded the secret image into the channel, which does not relate to color information. This idea effectively reduces the color distortion of color stego images\cite{r4}. More than this, the idea provides researchers with a new path: to use the characteristics of color spaces to improve the imperceptibility of image steganography based on deep learning and make it fit the visual characteristics of the human eyes. Nevertheless, the influence of color spaces characteristics still needs to be verified in different deep learning steganography.


According to the research results of traditional steganography, the embedding domain will have a significant impact on the steganographic visual effect\cite{r5}\cite{r6}\cite{r7}. According to the difference of embedding domain, image steganography can be divided into two categories: The first type is the spatial domain steganography, which directly modifies the pixels in the image space. The second type is frequency-domain steganography, which uses the frequency domain transformation method to convert images into frequency sub-bands and modify them. Contrary to image information in the spatial domain, the frequency domain sub-bands are information human eyes can not directly observe. Therefore, in the process of frequency domain steganography, the steganographic traces of stego images is invisible if the modification of frequency sub-band is controlled\cite{r8}. Unlike traditional steganography, image hiding steganography based on deep learning primarily hides secret data in the spatial domain of the cover image. The information in the spatial domain is directly visible to the human eyes. Therefore, if modified this information directly, it is easy to create a color distraction or artificial texture traces. Combined with the research results of traditional image steganography, we believe that image hiding steganography based on frequency domain will show a better imperceptibility through the learning of the deep neural network.

Based on the above inference, we try to embed a secret image into the frequency domain of a cover image to obtain better image quality. On this basis, the influence of different frequency sub-bands and color spaces on frequency domain steganography based on deep learning will be further discussed. In terms of frequency sub-bands, the characteristics of some frequency domain transform methods(such as DWT) are consistent with the characteristics of the human visual system (HVS)\cite{r9}. Hence Some traditional frequency domain steganography further improves the imperceptibility by using different frequency components and characteristics of HVS\cite{r10}. Similarly, we can further improve the imperceptibility of steganography by using the characteristics of image frequency sub-bands. In terms of the characteristics of color spaces, different color spaces will express image information in different ways. Hence they will produce different effects on steganography\cite{r11}. Moreover, each channel also has different visual characteristics in the same color space. Therefore, controlling each channel for steganography will also produce different steganography effects \cite{r12}. According to the characteristics of color spaces and the degree of image distortion, this paper selects the channel with the best steganographic effect as the embedding channel to further improve the imperceptibility of the model.

At present, image hiding steganography faces the problems of color distortion and artificial texture traces. According to the characteristics of color spaces and frequency sub-bands, a color image hiding steganography model based on frequency sub-band selection is proposed in this paper. The model is based on the deep neural network of encoder-decoder structure. Firstly, DWT is used to transform the secret image and the B channel of the cover image into the frequency domain. Then to reduce the color distortion and texture modification trace of the stego image, the high-frequency sub-band of the cover image is selected as the embedding domain. In order to keep a good balance between capacity and efficiency, we choose a gray image as the secret image and a color image as the cover image. The experimental results show that compared with other deep learning image steganography models based on spatial domain, the proposed model significantly improves image quality, reduces color distortion and artificial texture traces. The imperceptibility of the model is significantly improved. In summary, the main contributions of this paper are: 
\begin{enumerate}
 \item We propose a steganographic model in the frequency domain. Our model has better imperceptibility than the same category color image steganographic model in the spatial domain.
 \item	We analyze the influence of color spaces characteristics on the proposed model, find the most suitable embedding channel according to the characteristics of HVS and the principle of minimizing distortion.
 \item	Combined with the channel selection, we add the frequency sub-band selection module in the proposed model. This module selects the high-frequency sub-band as the embedding domain, which effectively improves the imperceptibility. Moreover, the relationship between frequency sub-bands characteristics and the steganographic effect is discussed.
 \item	To better compare the color difference between the original image and the stego image, we propose a new color image quality metric called CL-PSNR. CL-PSNR takes the color difference into account, which is more in line with the feeling of HVS to color images.
\end{enumerate}
The rest of the paper is organized as follows: In Section 2, we discuss the existing research results in related directions. In Section 3, we describe the structure of the model, the discussion of color spaces and imperceptibility, the selection of frequency sub-band, and the new color image quality metric. In Section 4, we present and analyze the experimental results. In Section 5, we conclude this work.

\section{Related Work}

Steganography is a technology of hiding information, which originated in ancient Greece and has a long history. The proposed steganography is the image hiding steganography based on deep learning, which means the steganography is based on deep learning knowledge and hiding the image with the image. Hence this section will introduce the development of image steganography based on deep learning. The representative achievements and problems of image hiding steganography will be discussed in detail. Moreover, the traditional image steganography will be used as a reference to improve imperceptibility.

\subsection{Image Steganography Based on Deep Learning}

In recent years, many image steganography methods based on deep learning have been proposed. ASDL-GAN was one of those earlier methods\cite{r13}. Although ASDL-GAN does not directly participate in the embedding and extraction process, its results inspire more scholars to improve the performance of steganography by using deep learning. Zhu et al. proposed Hidden, which first used a neural network for end-to-end training and has stronger robustness than traditional methods\cite{r14}. Hayes et al. proposed a HayesGAN to use an encoder network to hide text information in the image \cite{r15}. Tancik of Berkeley University proposed a StegaStamp framework based on the U-net structure \cite{r16}. They used a series of image enhancement methods to make the model robust in the real world. Huang et al. introduced the adversarial strategy in model training to better resist steganalysis\cite{r17}. Compared with the mature and stable traditional image steganography, the advantage of deep learning image steganography is to use the deep neural network to learn the steganographic method independently. Traditional image steganography techniques rely on artificial design, so it is difficult to improve the performance further. Therefore, the introduction of deep learning broke the deadlock. Moreover, with the development of deep learning technology, image steganography based on deep learning will have more space to develop.

\subsection{image hiding steganography}

In the early stage of deep learning image steganography, most secret information is a string or a bitstream, and the steganographic capacity is small (mostly 0.2bpp-4bpp)\cite{r2}. Fortunately, the image hiding steganography appeared. Image hiding steganography based on deep learning uses the experience of other deep learning steganography. It uses the image to hide the image, which significantly increases the capacity. The deep steganographic model proposed by Baluja is the first model using two same-size images\cite{r3}. It extracts the secret image features through a preprocessing network, then hides a secret image into a cover image by a hidden network, and finally extracts it by an extraction network. This model significantly increases capacity (24bpp). After that, Baluja tried to embed two secret images into another image and confuse the specific information of secret images\cite{r18}. This method improves capacity and enhances the security of secret information. Wu et al. proposed the StegNet. It is based on a deep neural network, and it can also hide the same size image, which only modifies 0.76\% cover image\cite{r19}. Van et al. proposed a new training scheme, which modifies the error back propagation to speed up the training of the network\cite{r20}. The above results use color images as secret information. However, gray image is also common secret information in image hiding steganography. Compared with the color image, gray image only contains semantic information, and its capacity is only one-third of the color image. The characteristics make the gray image more suitable as secret information in specific tasks. Rehman et al. proposed a steganographic model based on the coder-decoder network\cite{r21}. The model can embed a gray image into an RGB color image. Moreover, a new loss function is introduced to ensure the end-to-end joint training of the encoder-decoder network. Experiments show that the model can still obtain good image quality with a large capacity(8bpp).

The image hiding steganography based on deep learning significantly improves the steganographic capacity, but its steganographic traces(color distortion or artificial texture traces) is visible to human eyes. Zhang et al. proposed ISGAN\cite{r4} to solve the problem of color distortion, which converts the RGB cover image into YUV color space, and then selects the Y channel of the cover image as the embedding domain to avoid the modification of color information. The model effectively reduces color distortion. However, slight brightness differences still exist. Hence there is still room for improvement of imperceptibility.

Most of the image steganography methods based on deep learning are operate in the spatial domain. Unlike this kind of steganography, traditional image steganography can be carried out in both spatial and frequency domains. One of the most typical spatial domain steganography is to replace the least significant bit (LSB) of pixels with secret information\cite{r22}.On the other hand, Outguess\cite{r23} and J-UNIWARD\cite{r24} are typical frequency domain steganography. In traditional image steganography, frequency domain steganography usually obtain better robustness and can resist the attack of steganalysis more effectively. Moreover, thanks to the characteristics of frequency transformation, some frequency domain steganography use the HVS characteristics to process frequency domain information to improve the visual quality of the generated image\cite{r9}. There is almost no deep learning image steganography using frequency domain to hide the image. Therefore, relying on the powerful learning ability of deep neural networks and advantages of the frequency domain, the performance of image steganography is expected to be further improved. Although image hiding steganography based on deep learning effectively improves capacity, there are still noticeable color distortion and artificial texture traces. Inspired by traditional frequency domain steganography, this paper attempts to establish a frequency domain color image steganography model based on deep learning. We use the characteristics of color spaces and frequency sub-bands to improve imperceptibility further. Our model shows significant imperceptibility when keeping a large capacity compared with other models.

\section{Methodology}

\begin{figure}[h]
\centering
\includegraphics[width=1.0\textwidth]{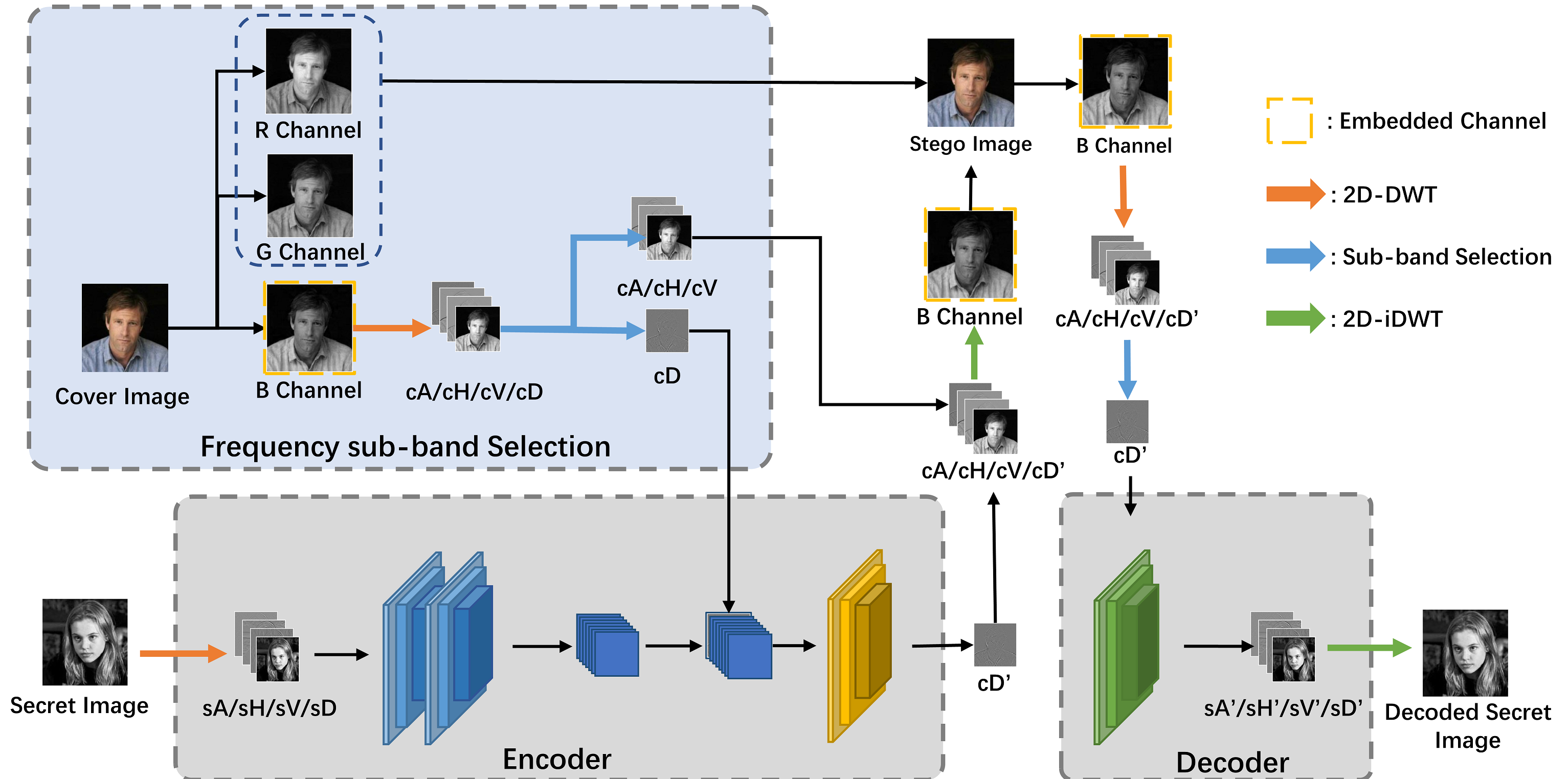}
\caption{\label{fig:3-1}The structure of the color image steganographic model based on frequency sub-band selection. The model consists of a frequency sub-band selection module, encoder network, and decoder network. The encoder network includes a pre-processed network for feature extraction of the secret image.}
\end{figure}

We aim to establish a color image steganographic model using a deep neural network based on the frequency domain. We explore methods to improve imperceptibility according to characteristics of color spaces and frequency sub-bands. The proposed model mainly considers how to improve imperceptibility in the case of large capacity (8bpp). Hence the robustness is not discussed in this paper. Our model is a fully convolutional encoder-decoder network. The key to improving imperceptibility is adding the frequency sub-band selection module. The model inputs are a color cover image and a secret gray image. Encoder and Decoder networks output a color stego image and a reconstructed gray secret image. The structure of the proposed model is shown in Figure \ref{fig:3-1}.

The frequency sub-band selection module pre-processes the cover image. The module selects the B channel as the embedding channel and separates the diagonal high-frequency sub-band (cD). The frequency sub-bands of secret image ([sA, sH, sV, sD]) and cD are the inputs of the encoder network. The encoder network first extracts the features from [sA, sH, sV, sD] and then encodes with cD to hide the secret image. For the extraction of secret information, the decoder network receives the diagonal high-frequency sub-band (cD') as input. Finally, the decoder network outputs the reconstructed secret image through the decoding and inverse DWT.

\subsection{Why B Channel}

In order to ensure that the stego image has a quality closer to the original cover image, we hope to retain the original image information as much as possible. Therefore, the proposed model only selects one channel of cover image as the embedding channel, which means only 1/3 of the cover image is modified. The data to process is reduced, and the learning efficiency of the network is improved.

\begin{figure}[t]
\centering
\includegraphics[width=1.0\textwidth]{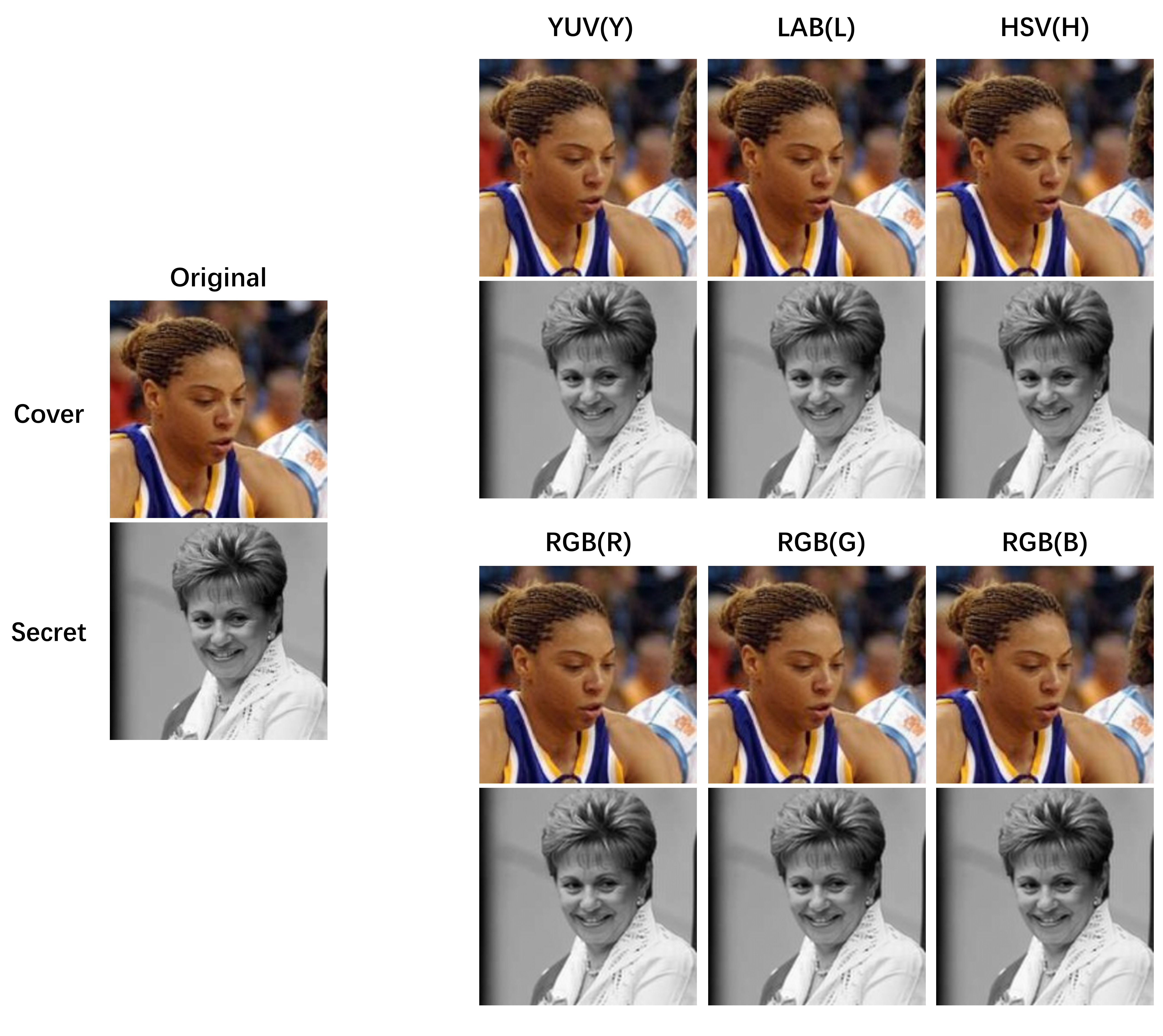}
\caption{\label{fig:3-2}A group of steganographic examples based on different embedding channels. The left column is the original cover image and the secret image (the second row) The others are the stego images and reconstructed secret images generated in the corresponding color space.}
\end{figure}

Considering the influence of color space characteristics, we need to find the optimal embedding channel to obtain the minimum image distortion and the best visual effect. Many steganographic models tried to improve the quality of the generated image by using channels independent of color information (for example, ISGAN uses the Y channel of YUV color space). Therefore, we select the embedding channel in several typical color spaces according to channel characteristics to view the image quality generated by the proposed model. The experimental results are shown Figure \ref{fig:3-2}.

We use the proposed model to hide information in different color spaces. We select the common nonlinear luminance/chroma color space (YUV, LAB), intensity/saturation/hue color space (HSV), and the original color space (RGB) and select a channel for embedding in each color space. Since modifying the brightness channel is more imperceptible, we give priority to the brightness channel for YUV, LAB, and HSV. The three channels of RGB color space represent three primary colors, so we hid the image in R, G, and B channels, respectively. As shown in Figure 4, all generated images are almost the same as the cover image. Therefore, the steganographic quality of the proposed model is hardly affected by the characteristics of color space.

It can be seen from previous studies that if the modification is visible, the sensitivity of vision to modification can be reduced by using a brightness channel as an embedding channel \cite{r4}. Vision is a subjective judgment. Hence we use the JND (the Just Notice Difference) to calculate the visual detection threshold. JND is the largest image distortion imperceptible to vision, which can judge whether the modification of the cover image reaches the threshold that human eyes can perceive. According to the calculation\cite{r25}, we can see that the JND threshold of the original cover image in Figure \ref{fig:3-2} is 7.46. However, the modification of stego images is far less than the threshold (the maximum modification is 1.30). Therefore, the modification caused by our model does not reach the range that human eyes can perceive. Moreover, in the following experimental part (4.2), we use the new metric proposed in this paper, CL-PSNR, closer to human visual perception to evaluate the quality of stego images. The results show that the CL-PSNR values of different color space experiments remain in the same range of good imperceptibility. Therefore, using other color spaces in our model can not effectively improve imperceptibility but will add unnecessary computational tasks and loss of accuracy (mostly from space transformation operations). Overall, our model uses the original RGB color space.

Although three stego images generated based on R, G, and B channels are almost the same in Figure \ref{fig:3-2}, we still select the embedding channel according to the principle of minimizing distortion. We calculate the error per pixel of each stego image, in which the image distortion from the model based on the B channel is the slightest(error per pixel is 0.66 BPP). Moreover, human vision is less sensitive to the modification of channel B than the other two color channels\cite{r26}. Hence B channel is also the best choice.

We select the B channel of RGB color space as the embedding channel based on the steganographic results in each color space. Based on the B channel, the stego image is visually close to the original image and has less distortion than that embedded in other channels.

\subsection{Selection of Frequency Sub-band}
After determining the embedding channel, the frequency sub-band selection module further selects a frequency sub-band as an embedding area. 

First, the B channel information in the spatial domain will be transformed into frequency domain information. At present, DCT(discrete cosine transform) and DWT are the most commonly used methods for frequency-domain transformation. Compared with DCT, DWT does not produce a blocking effect and can restore images more accurately. Moreover, DWT can also select different wavelet functions according by actual needs to decompose the image signal at different scales\cite{r27}. Therefore, to minimize the accuracy loss caused by frequency-domain transformation, we choose the DWT method to process B channel information. After experiments, we choose the dmey wavelet, which has a better network learning effect. In the complete steganographic process, the spatial-frequency domain transformation of images needs to be carried out twice, used for the data input of Encoder and Decoder. Similarly, the frequency-spatial domain transformation also needs to be carried out twice, used for the output of Encoder and Decoder.

In order to further improve the imperceptibility, we select only part of the frequency information of the cover image as the embedding domain instead of modifying all sub-bands. The secret information is a gray image of the same size as the cover image. Hence it is not easy to hide information and maintain good invisibility in a limited embedding domain. In order to ensure that the generated stego image is as close as possible to the original image, we hope that the embedding domain has as little impact on the whole image as possible. After DWT, We will get four frequency sub-bands, [cA, cH, cV, cD]. Where cA is a low-frequency sub-band,  corresponds to the region with smooth gray transformation and simple texture in the image. Besides, cH, cV, and cD are high-frequency sub-bands corresponding to the image region with intense gray transformation and complex texture. Although the four frequency sub-bands have the same dimension, most of the energy of the image will be concentrated on cA. On the contrary, cD only contains a small amount of detailed information. It means that the embedding area size provided by cD is the same as other sub-bands. However, with the same modification, its impact on the original image information is the smallest.

\begin{figure}[h]
\centering
\includegraphics[width=0.4\textwidth]{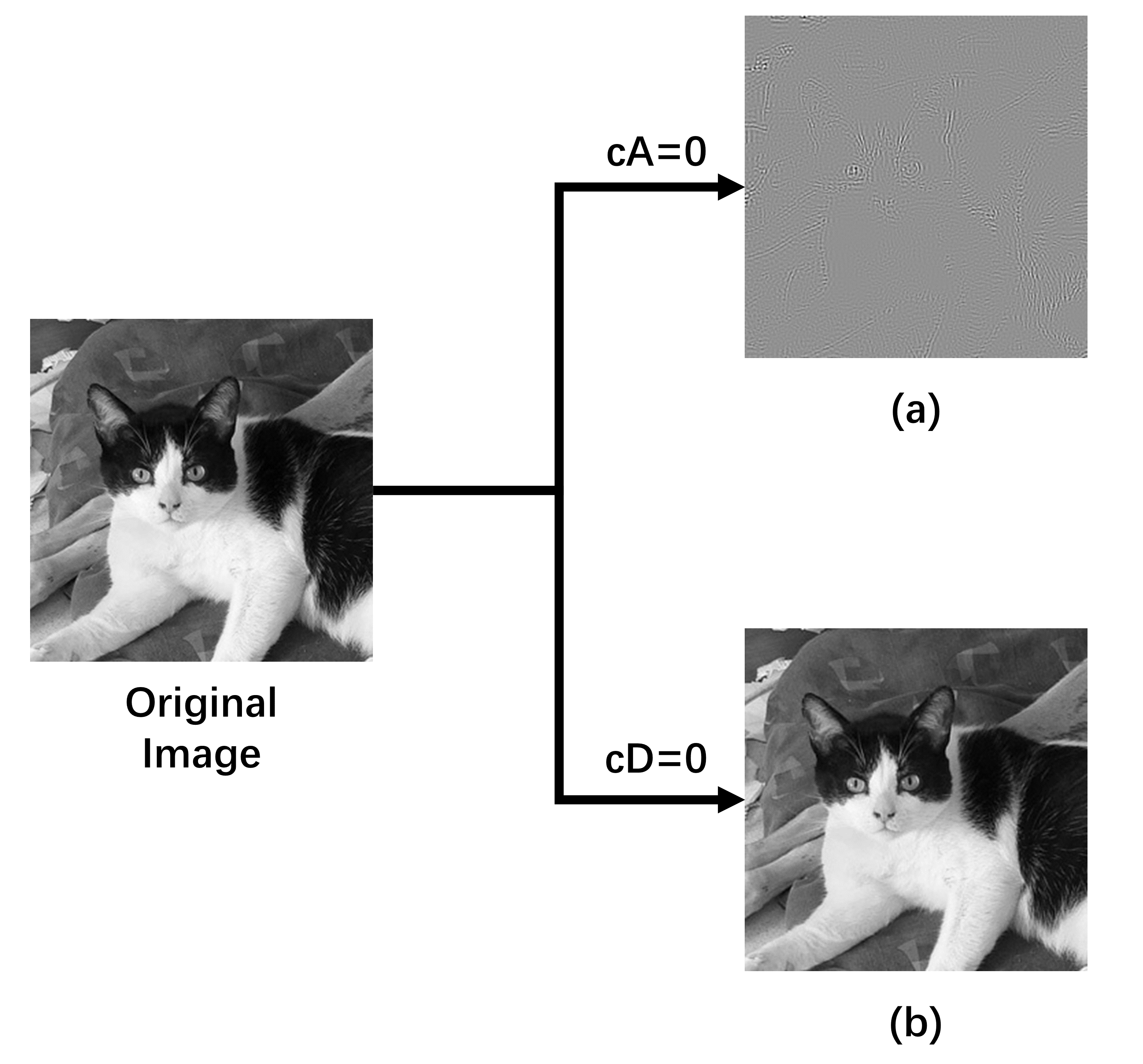}
\caption{\label{fig:3-4} (a) It is the result of setting the values of cA to 0. Almost all information in the original image is lost. (b)Tt is  the result obtained by setting the value of cD to 0, which is almost the same as the original image.}
\end{figure}

Figure \ref{fig:3-4} shows images after the same modification on different frequency sub-bands. We set all values in cA and cD to 0. The modification percentage of (a) (which means the percentage of pixels with value error larger than 5 bpp in the whole image) is as high as 99.39\%. In contrast, the modification percentage of (b) is only 10.97\%, and the error range of (b) is smaller than that of (a). Compared with the original image, the maximum pixel value error of (b) is 47.52, while the maximum pixel value error of (a) is as high as 255 (the maximum gray value). Incidentally, performing the same modification on cV and cH, the modification percentage is about 22\%. Although their visual results is better than (a), cD is still the best embedding area choice.

\begin{figure}[h]
\centering
\includegraphics[width=0.6\textwidth]{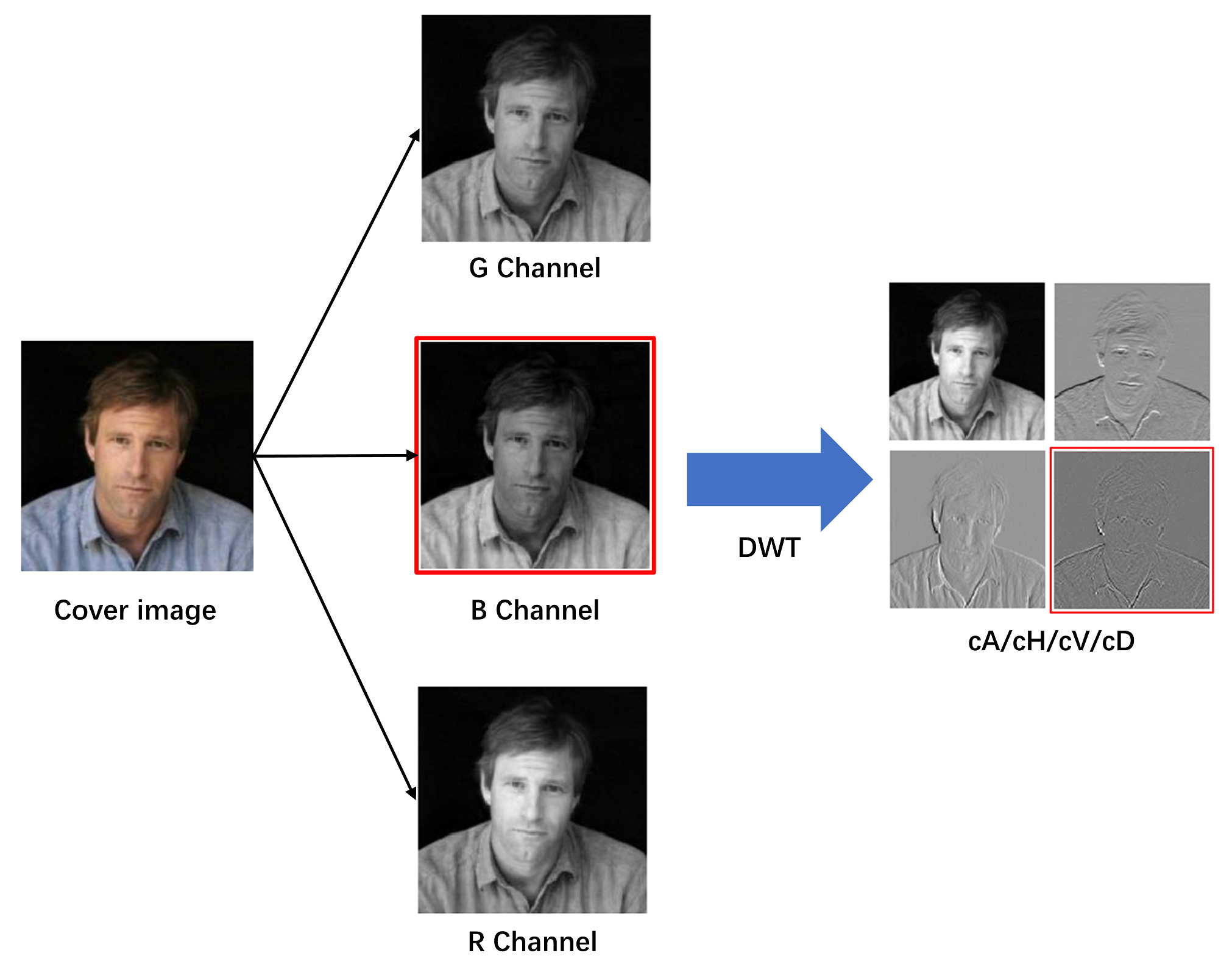}
\caption{\label{fig:3-5}The process of frequency sub-band selection module. Firstly, the B channel is separated, and then DWT operation is performed on the B channel to obtain frequency sub-bands. Finally, the model selects cD as the embedded area.}
\end{figure}

In addition to minimizing the modification range of the cover image, the characteristics of HVS are also an important reason we select cD. As shown Figure \ref{fig:3-5}, cA contains almost all the semantic information of the image. Low frequency often corresponds to gently transformed gray areas in the image, such as white walls, sky, etc. Therefore, it is easier for human eyes to detect the modification on cA, which does not fit the imperceptibility requirements of image steganography. Therefore, from the characteristics of HVS, cA is not suitable as an embedding domain. Unlike cA, other sub-bands contain detailed information such as the edge contour of the image. According to HVS characteristics, the vision is relatively insensitive to the modification of areas with complex texture, high contrast, or very bright/dark areas in the image \cite{r28}. Based on this property, we separate the diagonal high-frequency sub-band (cD) as the embedding domain. The cH and cV still contain low frequency information in one direction, so they are not considered embedding domains.

Figure \ref{fig:3-5} shows the process of the frequency sub-bands selection module. According to the influence of frequency sub-bands and the characteristics of HVS, the module selects cD as the embedding domain to reduce the impact of embedding on image quality. Therefore, the generated image has fewer texture traces and color distortion.

After embedding, the output of the encoder network will combine with other frequency sub-bands and be transformed into the spatial domain through inverse DWT. According to the embedding process, we could not measure the embedding position and modification of the stego image in the spatial domain. Therefore, we use the heat map to reflect the modification of the proposed model. Figure \ref{fig:3-7} shows a cover image, a stego image, and a heat map based on the difference between the two images.

\begin{figure}[h]
\centering
\includegraphics[width=0.6\textwidth]{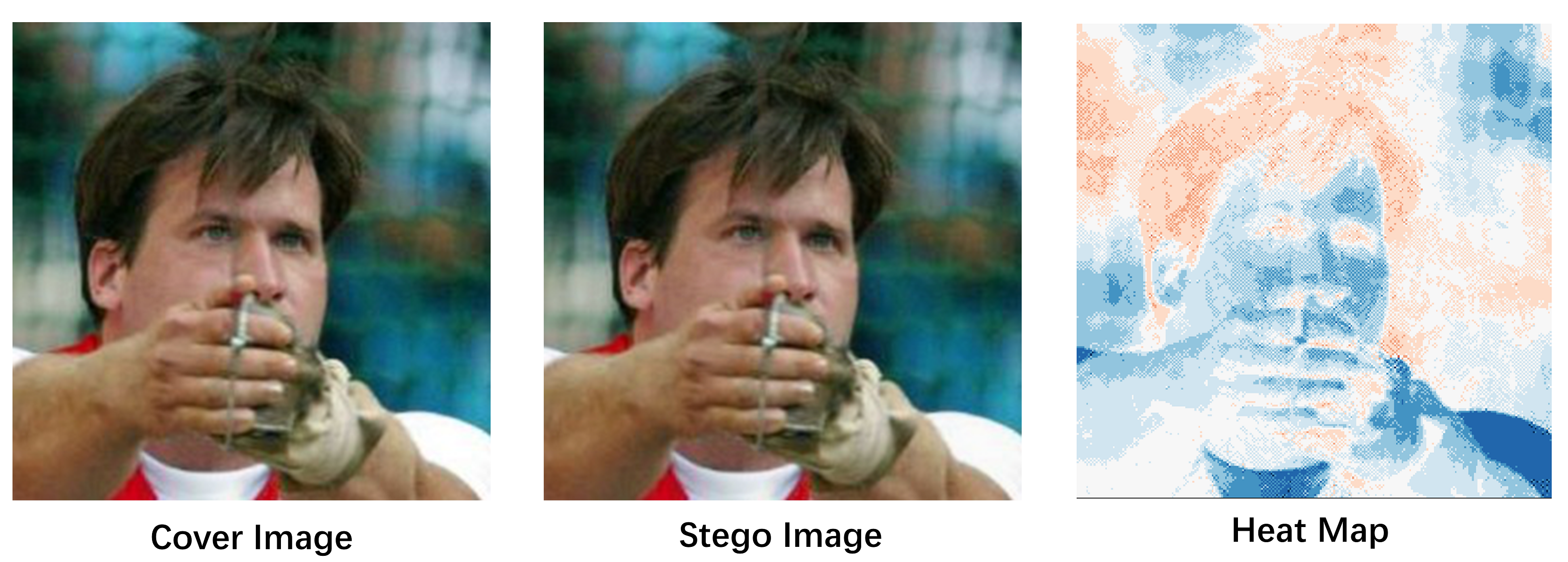}
\caption{\label{fig:3-6}The original image, the density image generated by the model, and the heat map based on the two image errors.}
\end{figure}

The heat map reflects the modification from steganography. In the heat map, red indicates that the pixel value of this position has increased, and blue indicates that the pixel value has decreased. The heat map has both red and blue areas, which means the model does not modify the cover image in only one direction. Since the color block in the heat map shows the outline of the cover image, we speculate that the network modification is determined by the data features of cD. Although almost all pixels in the heat map are colorful, the deviation of the overall pixel value is still within an acceptable range (in the example image, the maximum error is 14 bpp). As shown in Figure \ref{fig:3-6}, the stego image has a high image quality, which is almost the same as the cover image.

\subsection{Model Architecture}
After the pre-processing of the cover image, the model uses the encoder-decoder network to hide and extract the secret image. Encoder-decoder network is a common network structure in spatial domain steganography. We choose this network to explore whether it can learn the steganographic process in the frequency domain. Compared with other spatial domain steganographic models, the steganographic process is roughly the same, but the data transmitted in our model is frequency domain information.

\begin{figure}[h]
\centering
\includegraphics[width=1.0\textwidth]{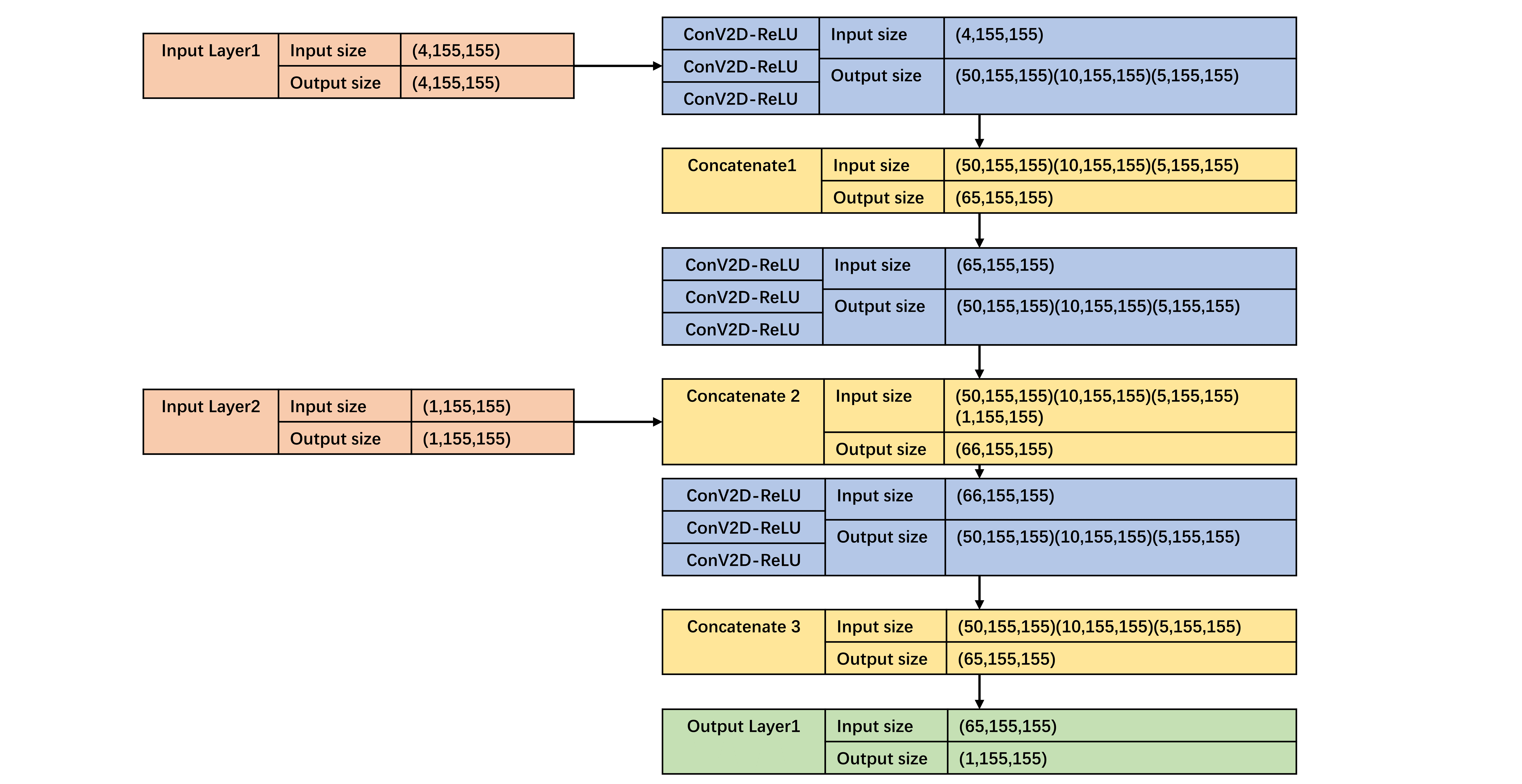}
\caption{\label{fig:3-7}Detailed configuration of the encoder network (using 250×250 images). The blue blocks are convolution layers, which are used to extract the features of secret image and embedding.}
\end{figure}

First, the encoder network embeds the secret image into cD. Encoder network is mainly composed of three groups of convolution layers, connection layers, and the input/output layer. The encoder network receives two inputs: the input layer 1 receives the frequency sub-bands [sA, sH, sV, sD], which are obtained after DWT of the secret image. The input layer 2 receives cD from the frequency sub-band selection module. The first two groups of convolution layers pre-process [sA, sH, sV, sD] to extract the data features of the secret image. The output feature maps are connected with cD, and the last group of convolution layers is used for coding.

The encoder network outputs the diagonal high-frequency sub-band (represented by cD'), which contains the frequency information of the secret image. Therefore, the stego image still needs to be obtained through the inverse operation of the frequency sub-band selection module.

The decoder network is mainly composed of a group of convolution layers, connection layers and the input/output layer. After receiving the stego image, the decoder network separates the cD' according to the frequency sub-band selection module. Input layer 3 receives cD'. Convolution layers perform a decoding operation to extract the frequency information of secret image (represented by [sA', sH', sV', sD']). Finally, the inverse DWT is used to transform the [sA ', sH', sV', sD'] into spatial domain information, that is, to obtain the reconstructed secret image.

\begin{figure}[h] 
\centering
\includegraphics[width=0.6\textwidth]{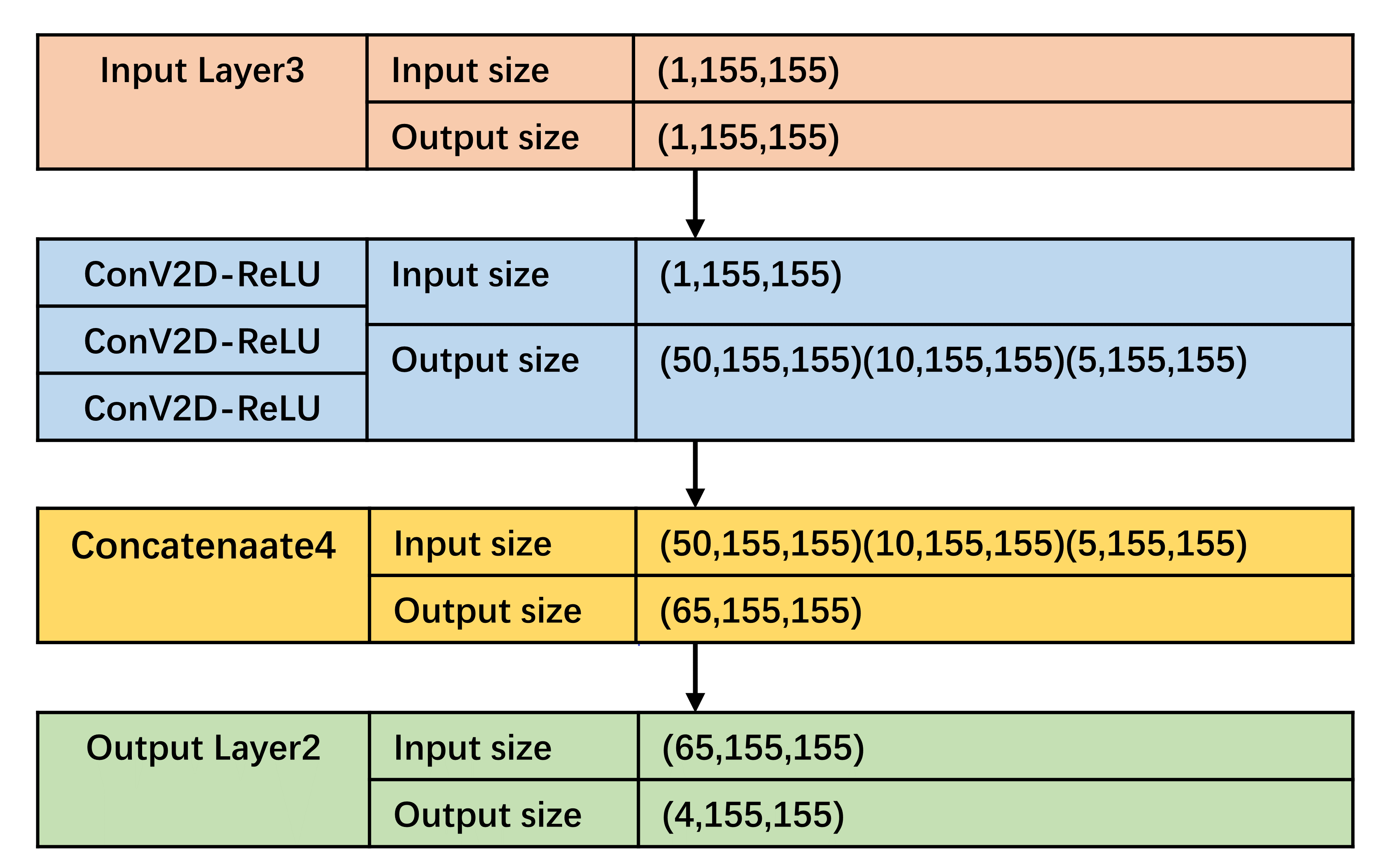}
\caption{\label{fig:3-8}Detailed configuration of the decoder network(using 250×250 images). The blue blocks are convolution layers, which are used to extract secret image.}
\end{figure}

In order to minimize the difference between the stego image and the original image, the secret image and the reconstructed secret image, the loss function we use is the MSE loss function calculated based on frequency domain:

\begin{equation} 
\centering
\text Loss(c,c',s,s') = \Vert{c-c'}\Vert^2+\beta \Vert{s-s'}\Vert^2
\end{equation}

In equation (1), $c$ and $c'$ represent the frequency domain information of the cover image and stego image, $s$ and $s'$ represent the frequency domain information of the original secret image and reconstructed secret image. In the encoder-decoder structure, image information is transmitted in the form of frequency information. Hence the loss function is also calculated in the frequency domain. The parameter $\beta$ is used to adjust the weight of the decoder network. Although the MSE loss function has achieved a good result in the proposed model, we still expect to design a loss function based on HVS characteristics in the future, which achieves a better imperceptibility.

We call the encoder-decoder network a basic model. In early experiments, we found that the steganographic effect of the basic model is better than that of the spatial domain steganographic model with the same structure(Table 2). This not only proves that the encoder-decoder network is also suitable for the deep learning steganography in the frequency domain, but also proves that the frequency domain is more conducive to network learning. Based on the basic model, we add a frequency sub-band selection module to form the final model, which further improves the imperceptibility of steganography.

\section{Experiments}

In this section, we designed a series of ablation experiments on color spaces and frequency sub-bands, determined the final model based on performances of imperceptibility. In the deep learning steganographic model of hiding a gray image with a color image, we choose Antique's model\cite{r21} and ISGAN\cite{r4} to compare with our final model. Finally, we verified the data generalization ability of the final model.

\subsection{Implementation Details}
\subsubsection{Dataset}
We mainly used the Labeled Faces in the Wild (LFW) dataset during our experiments. This dataset provides more than 13k face images which sizes are reshaped to 250×250. Due to multi posture, illumination, expression, age, occlusion, and other factors, even the photos of the same person are very different. We randomly selected 4K images as the training set and 200 images as the testing set. The training set and testing set are equally divided into cover images and secret images. Secret images are uniformly processed into gray images. In addition, ILSVRC and PASCAL VOC 2012 are also used for model training to verify the final model has a good generalization ability.

\subsubsection{Image Quality Evaluation Metric}
Firstly, we chose the widely used image quality evaluation metrics: PSNR (Peak Signal to Noise Ratio) and SSIM (Structural Similarity) to measure the quality of generated stego image and reconstructed secret image. 

During experiments, we found that PSNR sometimes can not directly reflect the perception of vision (especially when evaluating the quality of color images). In Figure 10, a and b are the stego images obtained by the proposed model in two different color spaces. For the observer, it is difficult to detect the difference between a and b, but the result difference of PSNR is more than 30. We believe that this is because PSNR is a metric designed for gray image quality evaluation. Therefore, when evaluating the quality of color image, PSNR sometimes gives results different from human eyes perception.

\begin{figure}[h]
\centering
\includegraphics[width=0.6\textwidth]{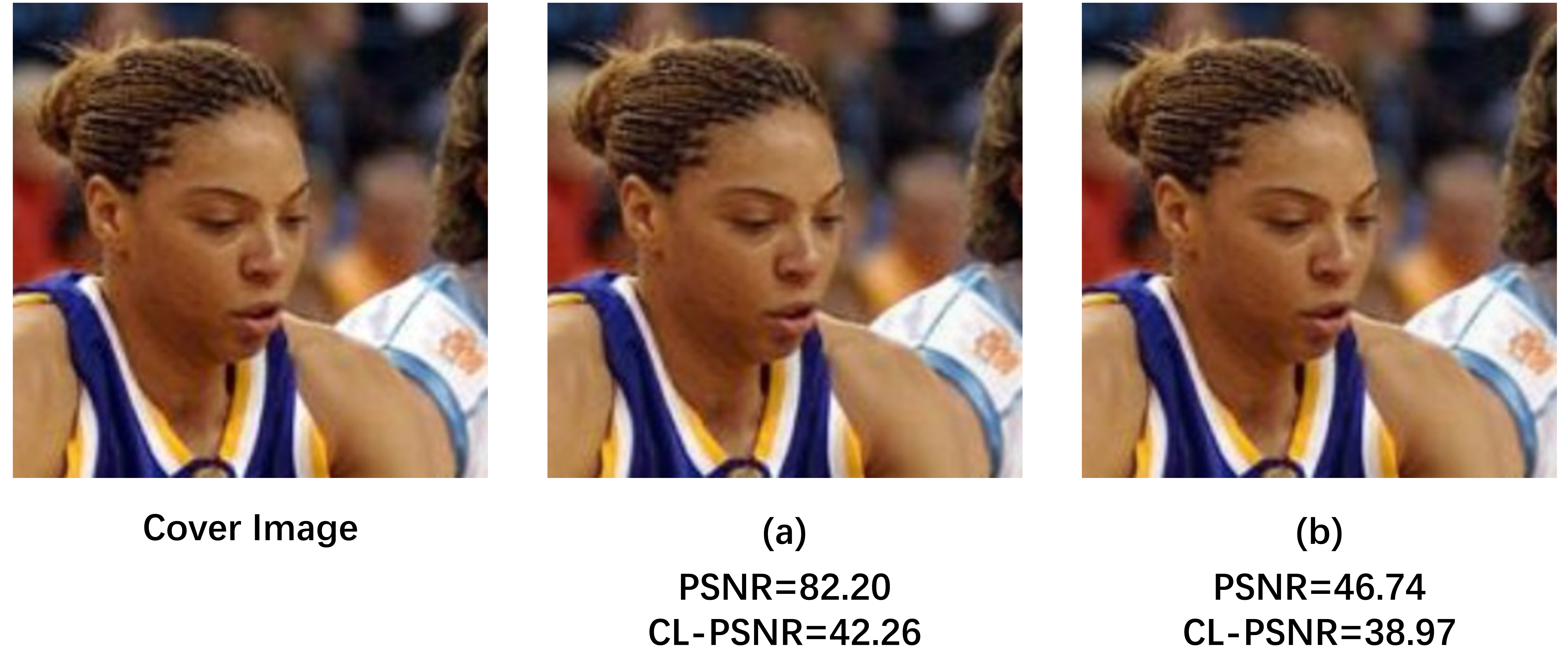}
\caption{\label{fig:4-1}Image a and b both have almost the same similarity with the cover image, but there is a large difference in their PSNR values. Compared with PSNR, CL-PSNR more fit with human visual perception.}
\end{figure}

For the limitations of PSNR, we introduce CIE color-difference formula\cite{r29} and propose a new image quality evaluation metric named CL-PSNR. We first calculate the color difference between generated image and the original image, then normalize it and calculate the average value to replace the MSE in the original PSNR equation, which is represented by CL-MSE.

\begin{equation}
\centering
\Delta E_{(x,y)} = \sqrt{ \left(\frac{\Delta L_{(x,y)}^{*} }{K_{L} S_{L}}\right)^2 + \left(\frac{\Delta C_{(x,y)}^{*} }{K_{C} S_{C}}\right)^2 + \left(\frac{\Delta H_{(x,y)}^{*} }{K_{H} S_{H}}\right)^2 }
\end{equation}

\begin{equation}
\centering
CL-MSE = \frac{\sum_{n=1}^{N} \frac{\Delta E_{n} }{MAX} }{N}
\end{equation}

Equation (2) calculates the CIE94 color-difference values of pixels $(L_{x}, a_{x}, b_{x})$ and $(L_{y}, a_{y}, b_{y})$, where $\Delta L_{*}$, $\Delta C_{AB}^{*}$, and $\Delta H_{AB}^{*}$ represent the lightness difference, saturation difference, and hue difference. $K_{L}$, $K_{C}$ and $K_{H}$ are parameter factors. $S_{L}$, $S_{C}$, and $S_{H}$ are correction coefficients of brightness, chroma, and hue. The default values for $K_{L}$ and $S_{L}$ are 1, and the values of $K_{C}$, $K_{H}$, $S_{C}$, and $S_{H}$ are adjusted according to different applications. In equation (3), $N$ means the total number of pixels. $MAX$ represents the maximum CIE color difference, rounded to 140.

\begin{equation}
\centering
CL-PSNR = 10 \times log_{10}\left( \frac{(2^{n}-1)^2}{CL-MSE} \right)
\end{equation}

In equation (4), we use CL-MSE to replace MSE in the original PSNR equation, so CL-PSNR will consider the influence of color difference.

Compared with PSNR, CL-PSNR is more fit with human visual perception. In order to further prove it, we select 100 pictures containing four different distortion types in TID2013 dataset\cite{r30} for verification.

\begin{figure}[t]
\centering
\includegraphics[width=0.5\textwidth]{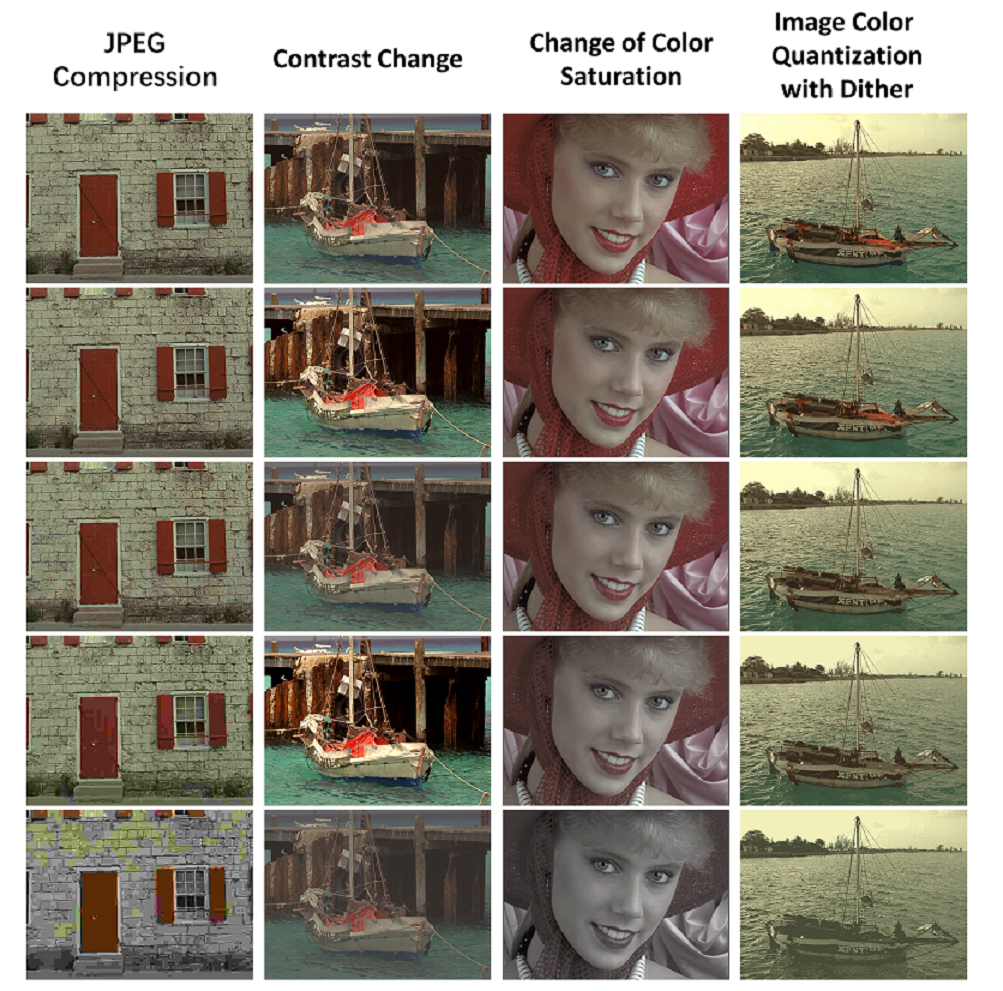}
\caption{\label{fig:4-2}Selected images in TID2013. The distortion types involved JPEG compression, contrast change, color saturation change, and image color quantization with dither.}
\end{figure}

\begin{figure}[h]
\centering
\includegraphics[width=0.5\textwidth]{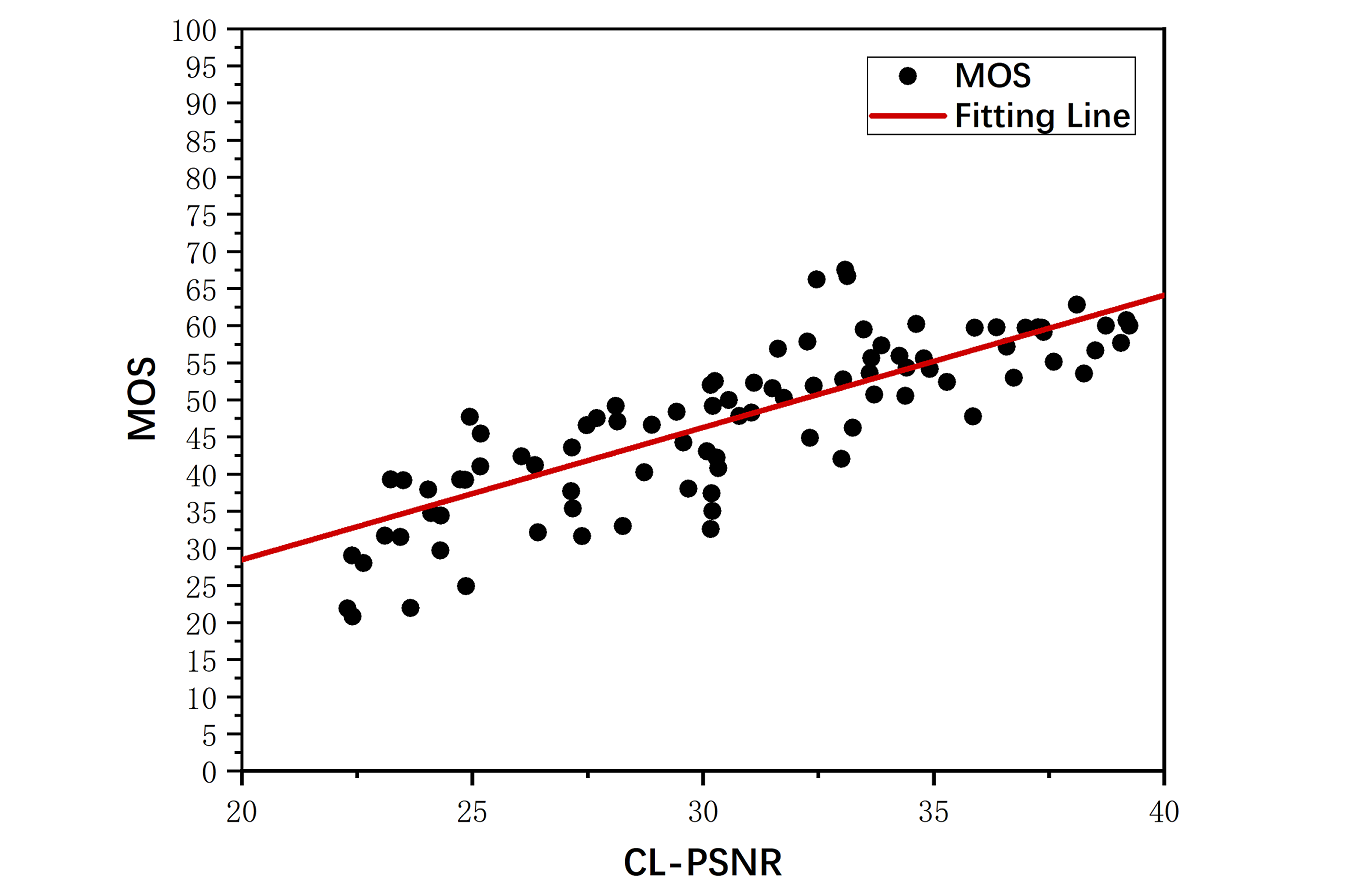}
\caption{\label{fig:4-3}CL-PSNR and MOS(Mean Opinion Score) distribution.}
\end{figure}

In Figure \ref{fig:4-3}, the vertical axis coordinate represents the Mean Opinion Score (MOS) given by the TID dataset, that is, the score for image quality obtained through subjective human eye experiment. We calculated the CL-PSNR values of sample images, represented them with points in the distribution diagram, and gave the fitting line. From the distribution trend, we can see that the MOS fraction increases with the increase of CL-PSNR, and CL-PSNR is linearly correlated with the MOS score. Therefore, CL-PSNR is suitable for the evaluation of color image distortion. We will add CL-PSNR as a new metric in experiments to evaluate the quality of the color stego image.

\subsubsection{Neural Network Training Settings}
We use Adam optimizer and train our model for 400 epochs. We set the initial learning rate to 0.001 and decreased to 0.0003 after 150 epochs. After 300 epochs, it decreases to 0.0001. The batch size is 16, and $\beta$ is 1 in the loss function. The model is implemented with TensorFlow, and an NVIDIA Tesla T4 GPU is used.
\begin{figure}[h]
\centering
\includegraphics[width=0.8\textwidth]{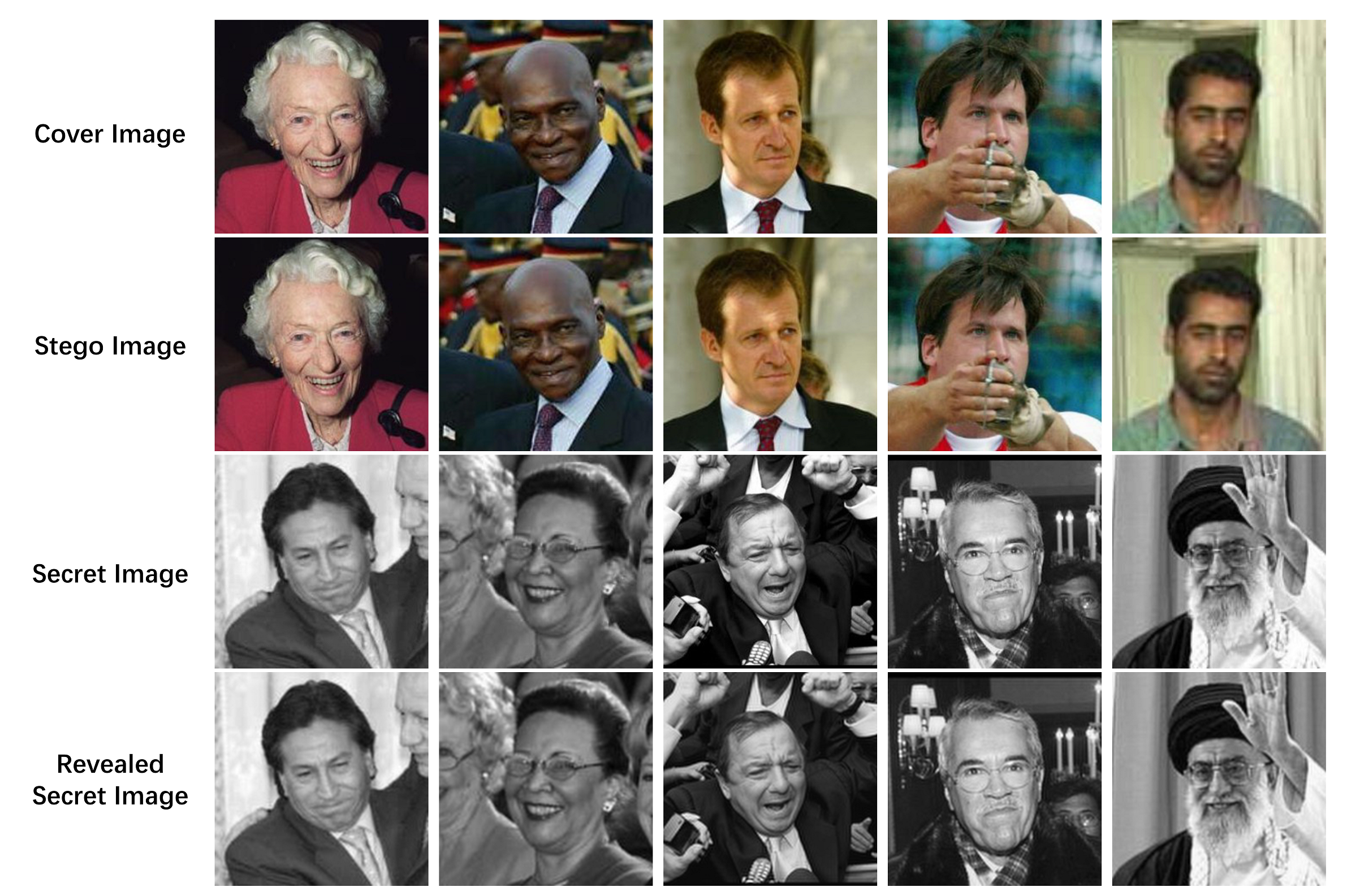}
\caption{\label{fig:4-4}Five groups of generated images by the final model. It is difficult for human eyes to distinguish the difference between original images and generated images.}
\end{figure}

Figure \ref{fig:4-4} shows the steganographic effect of the final models. The embedded secret information is a gray image of the same size as the cover image. Hence the capacity of the model is fixed at 8 bpp.

\subsection{Ablation Experiments}
We designed two ablation experiments to verify the effects of color spaces and frequency sub-bands characteristics on the proposed model. We determined the final model, which uses the cD of the B channel as the embedding domain.

\subsubsection{Color Spaces}
The following six experiments are designed to verify the superiority of the B channel as an embedding channel.

From the perspective of imperceptibility, in order to reduce the sensitivity of human eyes to steganographic traces, we select brightness channels for embedding in YUV, LAB, and HSV color spaces. The three channels of RGB color space represent three primary colors, and there is little difference in the sensitivity of human eyes to their modification. Therefore, the last three experiments are embedded in R, G, and B channels, respectively. The experimental results in Table 1 include the results of error per pixel (denoted by C\_Error per pixel and S\_Error per pixel), peak signal to noise ratio (denoted by C\_PSNR and S\_PSNR), and structural similarity (denoted by C\_SSIM and S\_SSIM) of stego images and reconstructed secret images. We also give CL-PSNR values of stego images. The steganographic performances in each color space are given in figure 4 of Section 3.1, which will not be repeated in this section.

\begin{table}[h]
\centering
  
\begin{tabular}{lllllll}
\hline
\begin{tabular}[c]{@{}l@{}}Color space\\ /Embedding \\ Channel\end{tabular} & \begin{tabular}[c]{@{}l@{}}C\_Error \\ per \\ pixel\end{tabular} & \begin{tabular}[c]{@{}l@{}}S\_Error\\ per\\ pixel\end{tabular} & \begin{tabular}[c]{@{}l@{}}C\_PSNR/\\ CL-PSNR\end{tabular} & S\_PSNR & C\_SSIM & S\_SSIM         \\ 
\hline
YUV/Y & 1.1529 & 3.5101 & 46.89/40.62 & 37.22 & 0.9919    & 0.9790 \\ 
\hline
LAB/L & 1.3090  & 3.3554 & 45.80/34.36 & 37.61 & 0.9895   & 0.9803\\ 
\hline
HSV/H & 0.9815 & 3.3464  & 48.29/40.75 & 37.63 & 0.9941   & 0.9803\\ 
\hline
\textbf{RGB/R} & \textbf{0.7} & \textbf{3.31} & \textbf{82.12/43.37} & \textbf{37.72} & \textbf{0.9971} & \textbf{0.9999} \\ 
\hline
\textbf{RGB/G} & \textbf{0.69} & \textbf{3.32} & \textbf{82.31/43.67} & \textbf{37.7} & \textbf{0.997} & \textbf{0.9999} \\ 
\hline
\textbf{\begin{tabular}[c]{@{}l@{}}RGB/B\\ (final model)\end{tabular}} & \textbf{0.66} & \textbf{3.3586} & \textbf{82.31/44.33} & \textbf{37.75} & \textbf{0.9975} & \textbf{0.9999} \\ 
\hline
\end{tabular}
\caption{\label{tab:1}Comparison of steganographic results based on different color spaces or channels}
\end{table}

The experimental results show that in the above color spaces, the PSNR of stego images from each experiment is greater than 40, and the SSIM value is close to or greater than 0.99. The PSNR value of secret images is greater than 37, and the SSIM value is close to or greater than 0.98. Combined with table 1 and Figure 4, we believe that in the above color spaces, the generated images have the quality close to original images, and our model is hardly affected by the characteristics of color spaces. We believe that this is because, after frequency sub-bands selection, the model further reduces the steganographic range of the cover image. Hence the modification from embedding does not reach the JND threshold of the cover image.

From the first three experiments and the last three experiments, it can be seen that the results of error per pixel, PSNR, CL-PSNR, and SSIM of steganography model based on RGB color space are better than other models. We think this is because the rounding operation of color space transformation may bring additional loss. Based on the principle of distortion minimization, we selected the B channel as the embedding channel of the proposed model. Compared with the modification of the other two color channels, human eyes are more insensitive to the modification of channel B. Moreover, the distortion under B channel is the smallest.

\subsubsection{Frequency sub-band selection}
In this section, we verified the advantages of frequency sub-bands selection. As shown in Table \ref{tab:2}, we compared the steganographic effects of models under different embedding domains based on the B channel. The embedding domain of the four experiments is the global spatial domain, the global frequency domain(cA+cH+cV+cD), the low frequency sub-band (cA), and the diagonal high-frequency sub-band (cD).
\begin{table}[h]
\centering

\begin{tabular}{lllllll}
\hline
Embedding Area & \begin{tabular}[c]{@{}l@{}}C\_Error per \\ pixel\end{tabular} & \begin{tabular}[c]{@{}l@{}}S\_Error per \\ pixel\end{tabular} & \begin{tabular}[c]{@{}l@{}}C\_PSNR \\ /CL\_PSNR\end{tabular} & S\_PSNR & C\_SSIM & S\_SSIM \\ \hline
Spatial Domain & 5.38 & 10.69 & 76.24/27.24  & 27.54 & 0.9225 & 0.7739 \\ 
\hline
\begin{tabular}[c]{@{}l@{}}Frequency domain\\ (basic model)\end{tabular} & 1.73  & 3.40 & 79.52/38.03 & 37.47   & 0.9923 & 0.9785\\ 
\hline
\begin{tabular}[c]{@{}l@{}}Low Frequency \\ sub-band\end{tabular} & 5.51 & 11.50  & 76.17/27.29 & 26.91 & 0.9211 & 0.7964 \\ 
\hline
\textbf{\begin{tabular}[c]{@{}l@{}}High Frequency \\ sub-band\\ (final model)\end{tabular}} & \textbf{0.66}  & \textbf{3.3586} & \textbf{82.31/44.33}  & \textbf{37.75} & \textbf{0.9975} & \textbf{0.9999} \\ 
\hline
\end{tabular}
\caption{\label{tab:2} Results of steganography based on different embedding domains}  
\end{table}

\begin{figure}[h]
\centering
\includegraphics[width=1.0\textwidth]{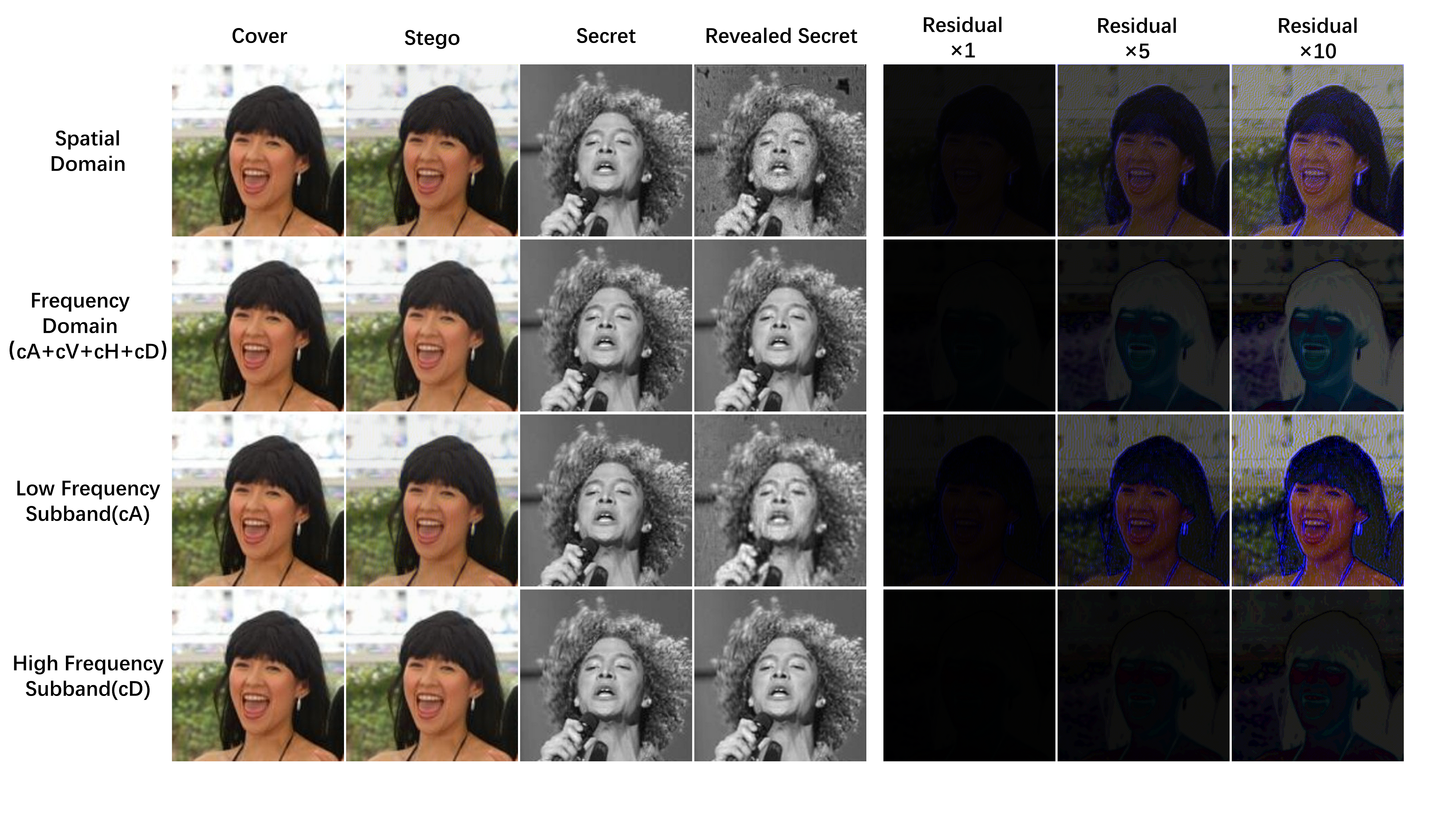}
\caption{\label{fig:4-5}The first and third columns are the original cover image and secret image, and other images are the generated images. The right side of each line shows the residual images of stego images.}
\end{figure}

According to Table \ref{tab:2} and Figure \ref{fig:4-5}, the secret image generated based on spatial domain has obvious distortion. In contrast, the distortion of the image embedded in the frequency domain is slighter and visually closer to the original image (the second line of Figure \ref{fig:4-5}). Therefore, the proposed model can obtain a better imperceptibility in the frequency domain. We believe that this is due to the different characteristics of spatial domain information and frequency domain information. Compared with spatial domain information, frequency domain information is the result of image decomposition from multiple different scales, and its characteristics are more conducive to network learning.

According to the experimental results of the last three groups in Table \ref{tab:2}, the C\_Error per pixel of the final model is only 0.66, the PSNR is as high as 82.31, and the SSIM of the stego image and the reconstructed secret image is almost saturated. In the residual images of Figure \ref{fig:4-5}, the steganographic trace generated by the final model is the smallest, so its stego image is closer to the original cover image. We attribute the advantage of the final model to the local modification of frequency domain and avoiding modifying cA, which has a large image energy concentration. When cD is selected for steganography, the dimension of its data accounts for only 1 / 4 of the overall frequency domain. And the data dimension is equal to other sub-bands, but it has less impact on the original image.

According to the above experimental results based on color spaces characteristics and frequency sub-bands characteristics, the module frequency sub-band selection of the final model selects the diagonal high-frequency sub-band of the B channel as the embedding domain.

\subsection{Comparison with Other Models}
In order to verify that the proposed model is superior to similar steganographic models in imperceptibility, we compare the steganographic effect of our model with that of Antique's model\cite{r21} and ISGAN\cite{r4}. Both of them are steganographic models based on a color image hiding a gray image, so they have the same steganographic capacity (8bpp) as the proposed model. The three models are trained with the same dataset.
\begin{table}[h]
\centering

\begin{tabular}{ccccc}
\hline
Model & C\_PSNR & S\_PSNR & C\_SSIM & S\_SSIM \\ 
\hline
Antique’s Model & 33.7 & 39.9 & 0.95 & 0.96 \\ 
\hline
ISGAN & 34.63 & 33.63 & 0.95 & 0.94 \\ 
\hline
Proposed Model & 82.31 & 37.75 & 0.99 & 0.99\\ 
\hline
\end{tabular}
\caption{\label{tab:3}Comparison of steganographic performance of Antique's model, ISGAN, and proposed model.}  
\end{table}

According to Table \ref{tab:3}, except S\_PSNR, other indicators of the final model are the highest, S\_SSIM and C\_SSIM of the final model reached 0.99, which is close to saturation. In addition, it is obvious from Figure 15 that the stego image generated by the final model is closer to the original image. And the distortion of the reconstructed secret image is also really difficult to detect. 

\begin{figure}[t]
\centering
\includegraphics[width=1.0\textwidth]{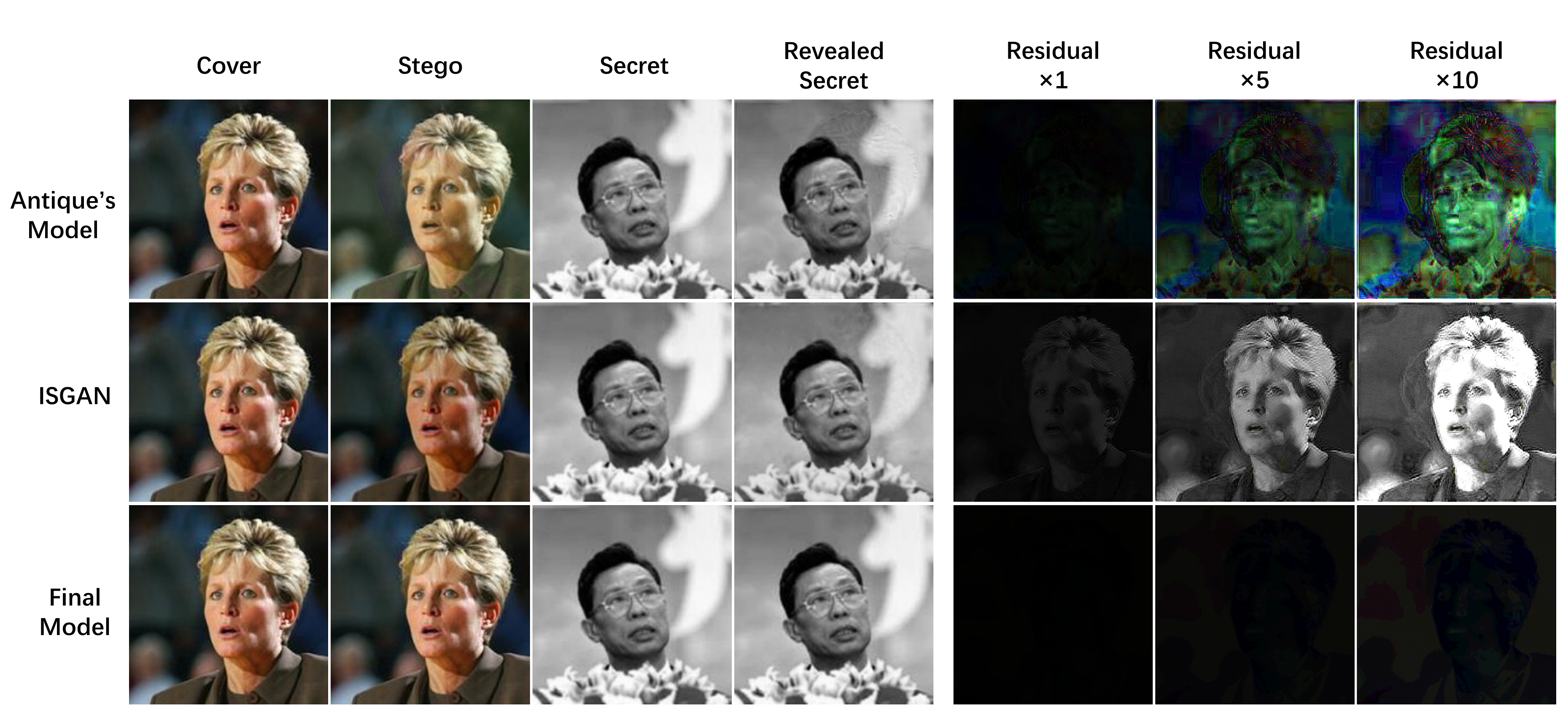}
\caption{\label{fig:4-6}The generated images and their residual images based on LFW dataset. The results show that compared with the models of ISGAN and Antique, the image generated by the final model is closer to the original image and has a good imperceptibility.}
\end{figure}

\begin{figure}[h]
\centering
\includegraphics[width=0.9\textwidth]{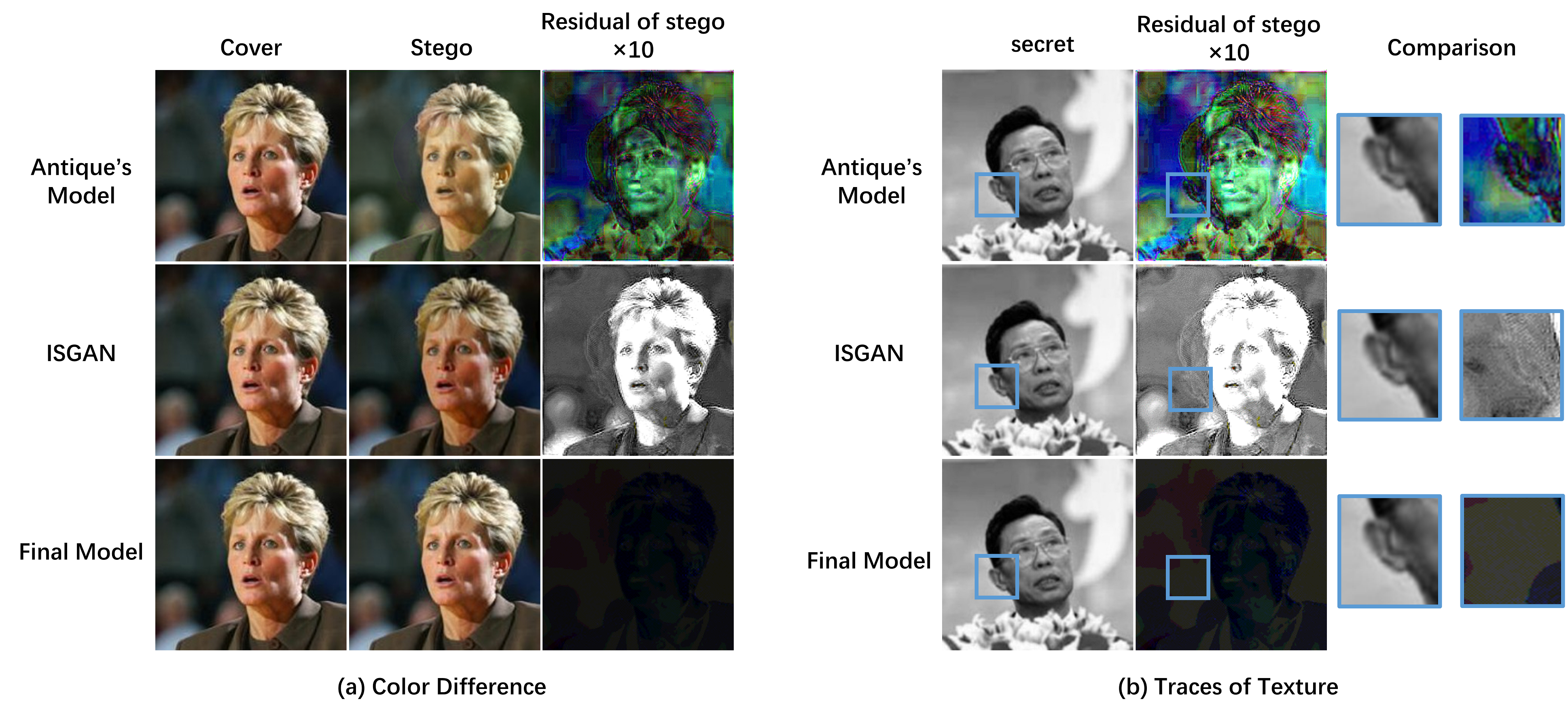}
\caption{\label{fig:4-7}(a) The color distortion between generated image and cover images; (b) The texture traces generated by steganography.}
\end{figure}

\subsubsection{Color Distortion}
Figure \ref{fig:4-6} (a) shows the color distortion of different models. It is obvious that the tone of the stego image generated by Antique's model (the second column) is yellower, the stego image of ISGAN is slightly darker, while the stego image of the final model is basically the original image, and the residual image does not show the obvious modification.

\subsubsection{Traces of Texture}
Figure \ref{fig:4-7} (b) shows the texture traces of different models. Compared with the color distortion, the texture traces of stego image are more difficult to detect by vision, but after enlarging the stego image or multiple residual processing with the original image, we can still find the texture outline of the secret image in the stego image of antique's model and ISGAN(we enhance the brightness of the residual image, so as to make the texture trace in the residual image more obvious). This means that the third party can obtain the semantic information of the secret image according to some simple image processing. Therefore, the security of the secret image is also difficult to guarantee. In contrast, the residual images of the final model still have no visible texture traces, which means our model has a better imperceptibility.

In summary, from the perspective of color distortion and texture traces, the proposed model has better performance in imperceptibility.

\subsection{Generalization Ability}
To further verify the generalization ability of the final model, we used the ILSVRC2012 and PASCAL\_VOC\_2012 dataset training model. 

\begin{figure}[h]
\centering
\includegraphics[width=1.0\textwidth]{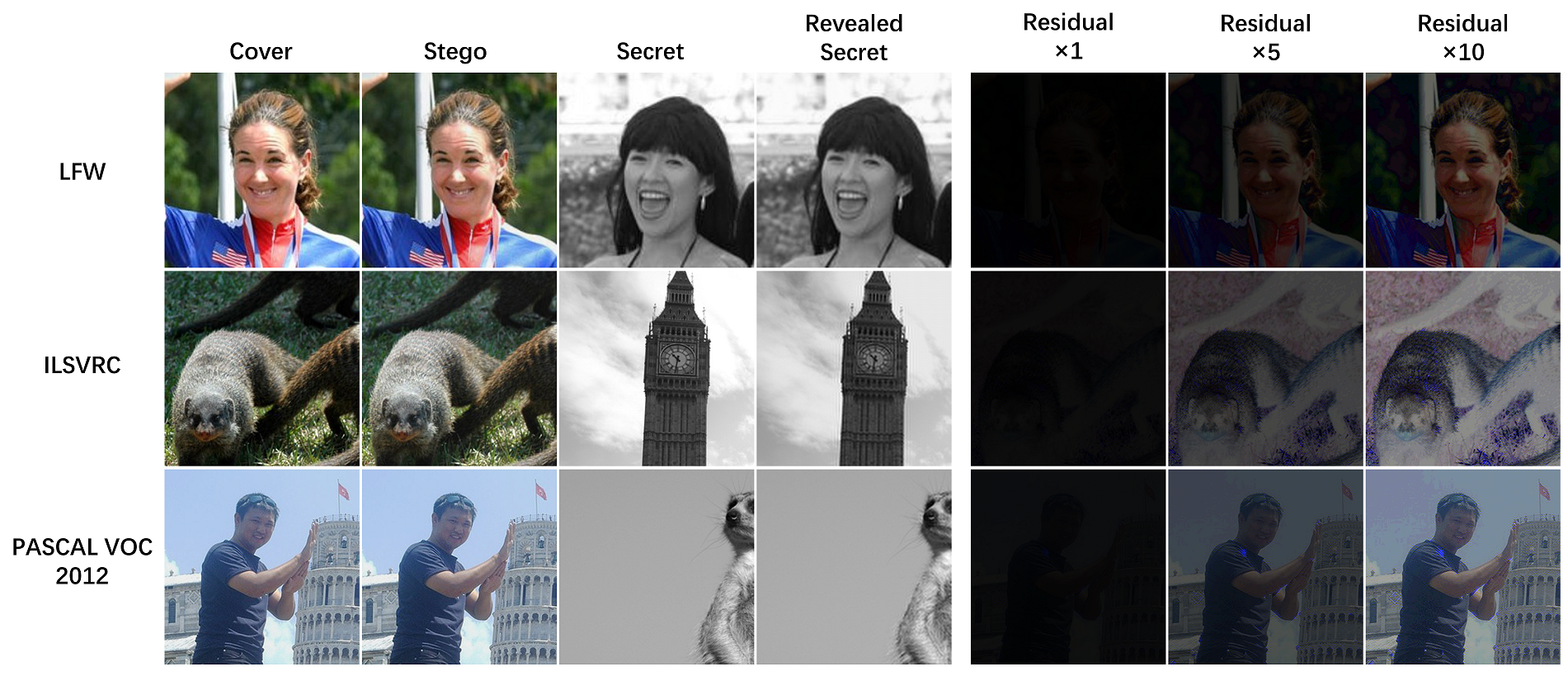}
\caption{\label{fig:4-8}Generated images based on different datasets. The final model can always generate images which close to original images.}
\end{figure}

\begin{table}[h]
\centering
\begin{tabular}{ccccccc}
\hline
Dataset& \begin{tabular}[c]{@{}c@{}}C\_Error \\ per \\ pixel\end{tabular} & \begin{tabular}[c]{@{}c@{}}S\_Error \\ per \\ pixel\end{tabular} & \begin{tabular}[c]{@{}c@{}}C\_PSNR\\ /CL-PSNR\end{tabular} & S\_PSNR & C\_SSIM & S\_SSIM \\ 
\hline
LFW& 0.66 & 3.3586 & 82.31/44.33  & 37.75 & 0.9975 & 0.9999  \\ 
\hline
ILSVRC2012 & 2.86 & 4.21 & 78.07/35.11 & 35.62 & 0.9914  & 0.966   \\ 
\hline
PASCAL\_VOC\_2012 & 2.67 & 6.02 & 78.30/35.49 & 33.05 & 0.9918 & 0.9466  \\ 
\hline
\end{tabular}
\caption{\label{tab:4} Training results of the final model based on different datasets}  
\end{table}
The data of the ILSVRC2012 dataset comes from the ImageNet dataset. PASCAL\_VOC\_2012 dataset is commonly used for target detection, including more than 20 object classes and more than 10 action classes. Compared with the LFW dataset used in previous experiments, the data characteristics of the two datasets are more complex, which is helpful for the generalization verification of the final model.

Although the performances of training on ILSVRC2012 and Pascal\_VOC\_2012 are slightly lower than that of training on LFW, but its distortion is still within an acceptable range. In addition, all SSIM values of the three results are more than 0.99, which is close to saturation. The last three columns of images in Figure \ref{fig:4-8} are residual images of the original cover image and the stego image. When the residual is expanded to 5 times, the outline of the stego image appears, and when it is expanded to 10 times, the outline is clearer. This means that the final model can still show good imperceptibility on different datasets, and the steganographic performance remains at the same level. Therefore, the proposed model has good generalization ability.

\section{Conclusion}

 In this paper, a color image steganographic model based on frequency sub-band selection is proposed. The model uses the frequency domain as the embedding domain. Moreover, we discuss and verify whether the characteristics of different color spaces and frequency sub-bands will affect the proposed model, which helps us determine the final model. Our model is an encoder-decoder structure model, so the research conclusion has reference significance for the same type of steganographic model and determined the final model. Experiments show that compared with other models, our model effectively reduces the color distortion and texture traces, which significantly improves the imperceptibility. At present, the stego image generated by our model is almost the same as the original cover image. However, there is still room for improvement in the quality of the reconstructed secret image. Moreover, our model has excellent performances in imperceptibility and capacity. Hence we will pay more attention to the improvement of its robustness in our future work.

\printbibliography{} 
\end{document}